\documentclass[aps,prl,twocolumn,showpacs,superscriptaddress,groupedaddress]{revtex4}
\usepackage{graphicx}  
\usepackage{dcolumn}   
\usepackage{bm}        
\usepackage{amssymb}   
 \pdfoutput=1

\begin{document}

\title{Another comment on claims of a transition to the ultimate regime}
\author{Erik Lindborg}
\affiliation{Department of Engineering Mechanics, KTH, Osquars backe 18, SE-100 44, Stockholm, Sweden}
\date{\today}

\maketitle

Zhu {\em et al.} \cite{Zhu} carried out DNS of 2D Rayleigh-B\'enard convection (RBC) up to Rayleigh number $ Ra = 10^{14} $ and reported evidence of a transition to the `ultimate regime' of heat transfer predicted by Kraichnan \cite{Kraichnan62} for 3D RBC, with Nusselt number dependence $ Nu \sim Ra^{1/2} (\ln(Ra))^{-3/2} $, instead of the classical prediction $ Nu \sim Ra^{1/3} $ \cite{Malkus54}.  A curve fit made by Zhu {\em et al.} \cite{Zhu}, indicated that $ Nu \sim Ra^{0.35} $ for the four data points at $ Ra \in [10^{13}, 10^{14}] $, which they interpreted as evidence of a transition to the ultimate regime, despite the fact that the fit is much closer to the classical prediction.  Doering {\em et al.} \cite{Doering} analysed the results of \cite{Zhu} and concluded that they should be interpreted as evidence supporting the classical prediction.  In a reply, 
Zhu {\em et al.} \cite{Zhu2} reported that they had carried out two more simulations at $ Ra > 10^{14} $ and presented a new curve fit  showing $ Nu \sim Ra^{0.357}$. They claimed that the new results constituted  `overwhelming evidence' of a transition and that they had `irrefutably settled the issue'. 

Lindborg \cite{Lindborg} estimated the convergence time scale for the mean kinetic energy in 2D simulations to more than hundred times larger than the time at which the highest Rayleigh number  simulations of \cite{Zhu} were ended,  and questioned the conclusions of  \cite{Zhu, Zhu2}. 
In a recent review paper, Lohse \& Shishkina \cite{Lohse} repeat the claim that Zhu {\em et al.} \cite{Zhu2} `irrefutably settled the issue', without referencing \cite{Lindborg}. They acknowledge that the mean kinetic energy was not converged but claim, without presenting any evidence, that the Nusselt number was converged. The author of this comment seriously questions the validity of this statement. A plot of the time evolution of the Nusselt number from the four simulations at $ Ra \in [10^{10}, \, 10^{11}] $ (number 7,8,9 and 10, listed in the supplementary material of \cite{Zhu}) sent to the author by Zhu and Lohse, shows that the Nusselt number in the simulation at $ Ra = 10^{11} $ was far from being converged at the nondimensional time, t=250, at which the simulation at $ Ra = 10^{14} $ was ended. Unfortunately, the plot cannot be shown, because Lohse and Zhu do not grant the author permission to publish it.  In recent 2D simulations carried out by He {\em et al.} \cite{He}, which are not referenced by Lohse \& Shishkina, the Nusselt number was calculated up to $ Ra^{13} $. Up to $ Ra^{11} $, the Nusselt number curve of \cite{He} falls on top of the curve of \cite{Zhu}, while it falls above at $ Ra \in [10^{11}, \, 10^{13}] $, where it conforms to
$ Nu \sim Ra^{1/3} $. Most likely, the reason behind this difference is that the simulations by \cite{He} were better converged than the simulations by \cite{Zhu, Zhu2}.  

Apart from the curve fit, Zhu {\em et al.} \cite{Zhu2} presented three other pieces of evidence in support of their claim. As one of these three pieces, they observed that the velocity profiles were logarithmic close to the centreline of the convection cell, seemingly in accordance with the assumption under which the prediction of the `ultimate regime' was derived by Kraichnan,  who assumed that the profiles conform to the classical boundary layer mean velocity profile 
\begin{equation}  \label{LOG} 
U(z) =  u_{\tau} \left ( \frac{1}{\kappa} \ln \left ( \frac{u_\tau z}{\nu} \right)  + B \right ) \, .
\end{equation} 
Here,  $ u_\tau $ is the friction velocity,  $ \kappa $ is the K\'arm\'an constant, $ \nu $ is the kinematic viscosity and $ B $ is a constant which is {\em Reynolds number independent}.  However, Zhu {\em et al.} found that the velocity profiles conform to (\ref{LOG}),  with a constant $ B $ that is {\em increasing} quite strongly with Reynolds number. As pointed out by \cite{Lindborg}, such an observation does not support the theory of \cite{Kraichnan62}. On the contrary, it invalidates the theory.  He {\em et al.} \cite{He} made the same observation and concluded that it was an indication that the asymptotic high $ Ra $ range had been reached, in which they observed $ Nu \sim Ra^{1/3} $. 
The  two other pieces of evidence given by \cite{Zhu2} are completely circumstantial and rest on a number of {\em ad hoc} assumptions regarding  the dynamics before and after an imagined transition to the ultimate regime. 

Based on a review of 2D simulations of RBC, Lindborg \cite{Lindborg}  concluded that the Nusselt number dependence, most likely, conforms to the classical prediction $ Nu \sim Ra^{1/3} $  in the limit of high $ Ra $. 
There is nothing  in the results of \cite{Zhu, Zhu2}  that contradicts this conclusion. The recent results of \cite{He} add further support to $ Nu \sim Ra^{1/3} $.

\end{document}